\documentclass[useAMS,usenatbib]{mn2e}
\usepackage{graphicx}
\usepackage{epstopdf}
\title[Statistical relationship between succeeding solar flares]
{Statistical relationship between the succeeding solar
flares detected by the RHESSI satellite}
\author[L. G. Bal\'azs et al.]{L. G. Bal\'azs$^{1,3}$\thanks{E-mail,
balazs@konkoly.hu}, N. Gyenge$^{2}$, M. B. Kors\'os$^{2}$, T. Baranyi$^{2}$, E. Forg\'acs-Dajka$^{3}$, \newauthor and I. Ballai$^{4}$
\\
$^{1}$Konkoly Observatory,Research Centre for Astronomy and Earth Sciences, Hungarian Academy of Sciences, \\  Konkoly-Thege M. \'ut 13-17, Budapest, 1121, Hungary\\
$^{2}$Heliophysical Observatory,Research Centre for Astronomy and Earth Sciences, Hungarian Academy of Sciences, \\
  Debrecen, P.O.Box 30, H-4010, Hungary \\
  $^{3}$E\"otv\"os University, P\'azm\'any P\'eter s\'et\'any 1/A,
Budapest, 1117, Hungary\\
$^{4}$Solar Physics and Space Plasmas Research Centre (SP2RC), University of Sheffield,\\
Hounsfield Road, Hicks Building, Sheffield S3 7RH, UK
}

\begin{document}
\date{}
\pagerange{\pageref{firstpage}--\pageref{lastpage}}
\pubyear{}
\maketitle
\label{firstpage}

\begin{abstract}
The Reuven Ramaty High Energy Solar Spectroscopic Imager (RHESSI) has 
observed more than 80,000 solar energetic events since its launch on February $12^{\rm th}$, 2002. 
Using this large sample of observed flares, we studied the spatio-temporal 
relationship between succeeding flares.  
Our results show that the statistical relationship between the temporal and spatial 
differences of succeeding flares can be described as a power law of the form $R(t)\sim t^p$ 
with $p=0.327\pm 0.007$. We discuss the possible interpretations of this result 
as a characteristic function of a supposed underlying physics.
Different scenarios are considered to explain this relation, including the case 
where the connectivity between succeeding events is realised through a shock 
wave in the post Sedov-Taylor phase or where the spatial and temporal relationship 
between flares is supposed to be provided by an expanding flare area in the 
sub-diffusive regime. Furthermore, we cannot exclude the possibility that 
the physical process behind the statistical relationship is the reordering of 
the magnetic field by the flare or it is due to some unknown processes.
\end{abstract}


\section{Introduction}
Sudden or explosive releases of energy are common phenomena in different cosmic objects 
occurring within wide range of energy release rate from as low as $10^{17}$ J sec$^{-1}$ in case of 
the Sun until as high as $10^{47}$ J sec$^{-1}$ in gamma ray bursts at cosmological distances. 
In the present context, the notion of sudden or explosive event means that the ratio of the 
released energy ($E$) compared to rate of its change ($dE/dt$) is significantly shorter that 
the dynamical time scale of the object, i.e. the ratio of the characteristic physical size ($R$) 
and the propagation speed ($v$) of any kind of disturbance initiated during the release 
(see, e.g. \citealt{b38}). A typical characteristic of these explosions is that the rising time 
of energy release is very short in comparison with the decay phase independently of the physics 
behind these explosions.

Our study will focus on sudden energy releases in the solar atmosphere called flares. 
The systematic multi-wavelength study of flares revealed many details that help us in understanding 
the physics behind these phenomena, however, there are still many unanswered questions concerning 
the true dynamics and energetics of flares (e.g. \citealt{b1, SM11, Fea11}).  It is widely 
accepted that magnetohydrodynamic (MHD) processes are responsible for producing a flare.
The energy released in a flare could originate from magnetic reconnection at the top of magnetic 
loop where the magnetic flux tubes take an X-type configuration. During reconnection events, the 
magnetic energy stored in magnetic field lines is released in a very localised way into thermal 
and kinetic energy. This energy is transported from the site of reconnection via, e.g. radiation, 
slow and fast MHD shock waves, accelerated particles, and high-speed collimated hot plasma flows (jets). 
Flares usually occur in the solar corona but part of the released energy is transported downward to the 
lower atmosphere where large-scale disturbances were observed travelling in a large distance away from 
the flaring sites as e.g. sunquakes observed in the photosphere \citep{ko98, b10} or Moreton waves 
in the chromosphere \citep{mo60}. The energy of flares can be within a range of 
$10^{17}-10^{26} \, J$ (\citealt{b7}).

Observations often reveal pairs of flares that occur with close temporal and/or spatial proximity 
and a physical connection between events was supposed to exist. Flares that occur because of common 
physical reason in different active regions are called sympathetic flares, and large-scale coronal 
structures are the most probable causes of their connection (e.g. \citealt {Moon02, b55}). 
Recently \citet {Liuea09} found that four flares and two fast coronal mass ejections (CMEs) occurred 
with a causal relationship in an active region within $\sim$1.5 hr time interval. They called these 
types of solar flares occurring in the same active region with a causal relationship successive flares. 
 \citet{Zuc09} also found a sequence of successive destabilisation of the magnetic field configuration 
starting with a filament eruption (relatively cool, dense object of chromospheric material suspended 
in the corona by magnetic fields) and ended in a large flare within $\sim$2 hrs; they referred to the 
process as domino effect. \citet{Jiang09} presented the evidence for occurrences of magnetic interactions 
between a jet, a filament and coronal loops during a complex event, in which two flares sequentially 
occurred at different positions of the same active region stating with 1 hr time difference and two 
associated CMEs. Recently \citet{b52} presented a study of a multiple flare activity containing three 
small flares, and a major eruptive flare over the period of two hours.
They concluded that the small precursor flares (preflares) indicated the localised magnetic 
reconnections associated with different evolutionary stages of the filament in the pre-eruption 
phase and these events play a crucial role in destabilising the filament leading to a large-scale 
eruption. The preflare activity occurs in the form of discrete, localised X-ray brightenings observed 
between 2 and 50 min before the impulsive phase of the flare and filament acceleration. \citet{Chifor07} 
claim that the X-ray precursors provide evidence for a tether-cutting mechanism initially manifested as 
localised magnetic reconnection being a common trigger for both flare emission and filament eruption.
\citet{Kim08} also demonstrated that a preflare eruption and the main flare have a causal relation 
because they are triggered by a sequential tether-cutting process.

Based on these observations and conclusions, it is natural to question whether there are a significant 
number of flares physically connected in such way that their relationship can be revealed with 
statistical methods. This question is extensively debated in the literature but the statistical results 
presented so far have left this question open. The used methods usually focus on the study of flare waiting 
times distribution (WTD, the distribution of times between events) which can provide information about 
whether flares are independent events, or not (e.g. \citealt {W00, Moon02}). The results suggests that 
determination of WTD gives varied results, suggesting that the observed distribution may depend on the 
particular active region, on time, and that it also may be influenced by event definition and selection 
procedures \citep {W09}. The solar flare sympathy is probably a statistically weak effect \citep {WC06}, 
but the successive flares probably do not occur randomly in time and the WTD are regulated by solar flare 
mechanisms \citep {Kubo08}.
We apply an alternative statistical method that takes into account both the spatial and temporal 
distributions of flares.

The observational data used in the present study are the flares appearing in the list provided by the 
RHESSI satellite between 2002 and 2010.
The paper is structured as follows: in Section \ref{redat} we present the observational set-up of our study 
and data used in the analysis. Section \ref{srel} deals with the statistical relationship between spatial 
and temporal differences of succeeding flares. The discussion on the possible physical interpretations of 
the statistical results is given in Section \ref{disc}. Section \ref{sum} presents the summary of the main 
results and the conclusions. The paper includes an Appendix containing a gallery of scatter plots for different time intervals displaying 
the statistical relationship discussed in the paper.

\section{Data used in the analysis} \label{redat}
\subsection{RHESSI flare list}

Since its launch, the RHESSI satellite \citep{b11} has observed more than 80,000 events in 9 different 
energy channels. These events are displayed in a table consisting of the main parameters of flares (time 
of explosions, durations, peak intensities, total counts during the outburst, energy channel of the maximal 
energy at which the flare is still measurable, location on the solar disc and quality flags). 
The $x,y$ positions of flares given in the
RHESSI table refer to the apparent observed position of an event on the solar disc \citep{b24}. 
The flare position is the average of the map peak locations in different time intervals for the 
flare measured in the energy band from 6 to 12 keV. Based on this position, the number of the 
associated active region is determined and inserted into the flare list. RHESSI flares are mostly 
microflares of GOES class A, B, or C; the most frequent type of flare being GOES class B. 
The thermal energy observed by RHESSI at the time of peak emission in 6-12 keV is in the range 
$10^{19}-10^{23}\, J$ and has a median value of $10^{21}\, J$ \citep{b15, H08}.

RHESSI microflares typically show elongated loop-like structures, which are interpreted as cooling 
post-flare loops. At first, the thermal HXR emission at the loop-top is seen in the lower-energy bands. 
Later, the footpoints become visible in the higher bands because of the energy deposition of the 
nonthermal electrons penetrating to the loop footpoints. The hot material at the footpoints evaporates 
from the chromosphere to the corona to fill up the loop \citep{b7, H08} which can be seen as thermal 
loop source in the RHESSI images. The distribution of HXR (4 - 10 keV) source heights was found to be 
well fitted by an exponential distribution with a scale height of $6.1 \pm 0.3 \times 10^6$ m. 
The minimum observable height due to partially occulted sources was found to be $5.7 \pm 0.3 \times 10^6$ m. 
in the solar corona \citep{b14}.

\subsection{Data processing and selection criteria}

\begin{table}
 \center
 \caption{The number of events in different energy channels in 
the studied time interval (2002-2010).}
 \begin{tabular}{@{}lrr@{}}
 \hline
  Energy channel[keV]&All events&Selected events\\
 \hline
 6 $-$ 12 & 38058 & 15132 \\
 12 $-$ 25 & 9601 & 3431 \\
 25 $-$ 50 & 902 & 294 \\
 50 $-$ 100 & 116 & 46 \\
 100 $-$ 300 & 53 & 26 \\
 300 $-$ 800 & 8   & 4 \\
\hline
\end{tabular}\label{chan}
\end{table}

Our aim is to study the spatial and temporal relationship between RHESSI flares, thus we have 
to determine the spatial and temporal difference between two consecutive flares. The temporal 
difference was calculated by using the peak time of the flares. To derive spatial distances [m] 
between the locations of flares, at first we had to convert the $x,y$ positions of flares [arcsec] 
to latitude and longitude [deg] of the Carrington heliographic coordinate system. 
During the orthographic projection transformation, we took into account the apparent variation of 
the solar rotation axis and radius because of orbital motion of the Earth. The height of flares should 
be also taken into account but the distance from the solar surface to the flares is not known. 
Therefore, we used the height of $6 \times 10^6$ m in each case, which is about that height where the flares are 
most frequently observed by RHESSI \citep{b14}. 
After calculating the heliographic coordinates of flares, we computed their estimated spatial 
distance from their successors on a sphere with a radius of $R_{\odot}+6 \times 10^6$ m.
Obviously, the assumption for the height of flares inserts some systematic error in calculating the 
heliographic coordinates and the error of the position of flares increases with the increasing 
distance from the solar disk centre. 
To exclude the cases with relatively large errors, we have omitted all the events whose longitude 
measured from the central meridian (LCM) is grater than $\pm 60^{\deg}$.  We also determined several 
selection criteria related to the active region (AR) where the flare occurred. 
At first, we excluded those events that were not associated with an identified AR in the RHESSI flare list. 
In the remaining cases, the number of AR and the heliographic coordinates of flares were compared with 
the position data of active regions available in the SOHO/MDI-Debrecen Data (SDD, 1996-2010) catalogue \citep{Gy11}. 
It was checked whether the position of flare and that of the AR in SDD can be matched with a tolerance 
limit of $\pm 10^{\deg}$ latitude and $\pm 20^{\deg}$ longitude. Finally, only those flares were 
selected which had the same identified active regions both in SDD and in RHESSI flare list.

Further, some events were disregarded because of uncertainties of the observation of position or peak time. 
Each RHESSI orbit provides approximately 1 hr of solar observations followed by about 40 minutes of 
time spent in eclipse. Solar observations are further reduced by passages through the South Atlantic 
Anomaly (SAA), during which the detector counts are not recorded due to the high flux of energetic 
particles  \citep{b15}. 
To avoid using flares with misidentified peak times, we disregarded events that occurred during 
passing into the Earth's shadow or during data gap, or which were connected to the entering phase of 
the satellite above the SAA. We also omitted from further considerations the events with invalid position data.

We also considered the energy channel of the maximal energy at which the flare is still measurable 
regardless of the energy channel of its successor. 
The flares in the energy channel 6-12 keV have enough number for statistical studies in different 
phases of the solar cycle, thus we have confined our analysis to these types of events.
Table \ref{chan} shows the number of the events in the flare list and the number of the events meeting 
our selection criteria in various energy channels in the studied time interval (2002-2010).

\section{Statistical relationship between succeeding flares}
\label{srel}
\begin{figure}
\begin{centering}
\hspace*{-9mm}
\includegraphics[width=84mm]{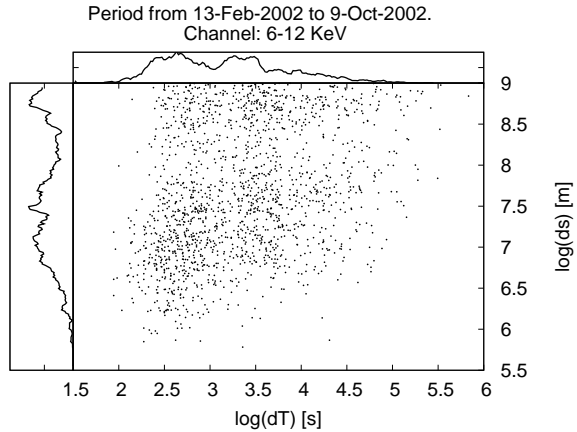}
\caption{The distribution of succeeding flares in the $\{\log_{10}(ds);\log_{10}(dT)\}$ plane. Note the two major populations 
of the events divided by a horizontal "valley" at about $10^8$ m. The rising trend in the lower 
population of events may indicate some relationship between the temporal and spatial differences of succeeding flares.}
\end{centering}
\label{example-fig}
\end{figure}

In order to study the statistical relationship between succeeding flares, we computed their spatial ({\it ds\/}) and temporal ({\it dT\/}) difference, i.e. the spatial and temporal distance of succeeding flares.  
We grouped the flare events according to the phase of the solar cycle to ensure equal number of considered cases. 
In this way, we selected  6 sub samples in the 6-12 keV energy channel.

At first, the {\it ds\/} and {\it dT\/} values were displayed in scatter plots based on the flares 
in the 6-12 keV energy channel without selecting out the succeeding flares in different active regions 
as it can be seen in Figure 1.
A correlation between the measured quantities for succeeding flares would mean some sort of casual 
relationship between them.
In this figure, the data points populate two major branches divided by a "valley" at about $ds=10^8$ m. 
The points above $10^8$ m do not show any relationship with the $dT$ time difference, i.e. there is no 
relationship between the flare and its successor in these cases.

On the contrary, the points below the "valley" populate a ridge having a well-defined rising tendency 
with $ds$ as the time difference is increasing between two succeeding flare events. This rising tendency 
might be interpreted as a possible causal relationship between succeeding flares.

Figure 1 also shows vertical strips containing more and less number of points in reflecting  the orbital 
motion of the RHESSI satellite around the Earth with maximum of points at about the orbital period of 
RHESSI (96 min) and with minimum of points at about 38 min because of the time spent in the Earth's shadow.

\begin{figure}
\includegraphics[width=84mm]{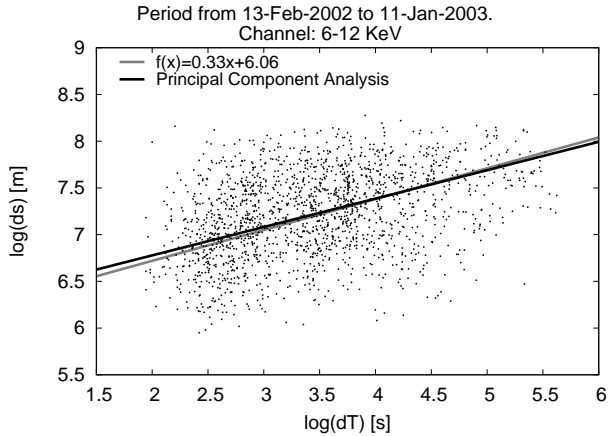}
\caption{Distribution of flares in the 6-12 keV channel and their successors 
in the $\{\log_{10}(ds);\log_{10}(dT)\}$ plane. 
The black solid line indicates the actual result of the PCA regression and the grey 
line is determined by the average parameters of Table \ref{param}.} \label{6_12-fig}
\end{figure}

Considering only the succeeding flares in the same active region the population of points above $ds=10^8$ m 
disappears and only the cases representing the rising trend remain (see Fig \ref{6_12-fig}).

To characterize the rising trend in the $\log_{10}(ds) \-\log_{10}(dT)$ diagram, we assume that the measured 
quantities can be written as  $\log_{10}(ds)=\log_{10}(R)+ \varepsilon_s$ and $\log_{10}(dT)=\log_{10}(t)+\varepsilon_t$ 
where
$R$ and $t$ are the true distance and time differences between two flares and 
$\varepsilon_s$, $\varepsilon_t$ represent noise terms. 
We represent the rising trend as a $\log_{10}(R)=p \log_{10}(t)+b$ linear relationship corresponding 
to an equation of the form
\begin{equation}\label{ST}
  R(t)=A t^p.
\end{equation}
The supposed relationship between the observed variables, $ds$,
$dT$, can be expressed by a linear regression model between the logarithmic variables:

\begin{equation}\label{eqds}
\log_{10} (ds)=a \log_{10} (t) + b + \varepsilon_s
\end{equation}
\begin{equation}\label{eqdT}
\log_{10} (dT)=\log_{10} (t) + \varepsilon_t
\end{equation}
\noindent where $a=p$ and $b=\log_{10}(A)$. The default procedure for the verification of the 
factor model is usually the principal component analysis (PCA) (see e.g. \citealt{Cadavid08}) 
performed on the covariance matrix of the $\log_{10}(ds)$, $\log_{10}(dT)$ variables.
Performing the PCA and keeping the first PC we obtain a variable running along the maximal 
variance direction of the points in the $\{\log_{10}(ds);\log_{10}(dT)\} $ plane. 
The regression line obtained in this way minimizes the sum of quadratic distances of the 
data points to a line. 
Therefore, this kind of regression is also called orthogonal and it gives the values of parameters $a$ and $b$ in the above 
system of equations.
The obtained parameters are summarized in the Table
\ref{param}. The errors in our estimations were calculated by means of the orthogonal 
regression (see, e.g. \citealt{b31}).

\begin{figure}
\includegraphics[width=84mm]{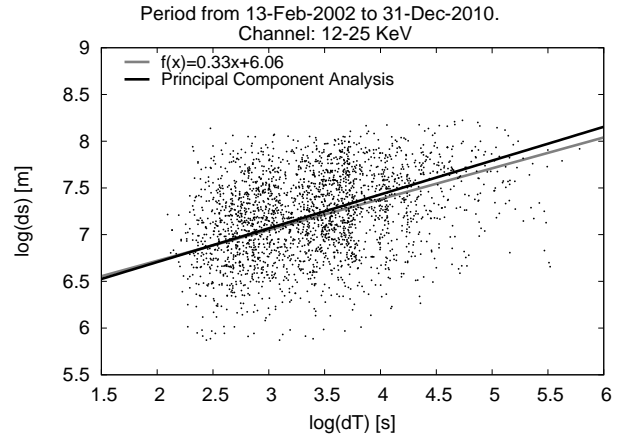}
\caption{Distribution of succeeding flares in the $\{\log_{10}(ds);\log_{10}(dT)\}$ plane 
in the 12-25 keV channel. The meaning of the lines is the same as in Figure \ref{6_12-fig}.} \label{12_25-fig}
\end{figure}

\begin{table}
 \center
 \caption{Parameters in Equation (\ref{eqds})-(\ref{eqdT}) obtained from the PCA regression 
for the flares in the 6-12 keV energy channel.}
 \begin{tabular}{llrrrrr}
 \hline
 Start & End & a & b & St.Dev. & St.Dev.\\
 date & date &  &  & (a) & (b)\\
 \hline
 2002-02-13 & 2003-01-11 & 0.304 & 6.170 & 0.017 & 0.087\\
 2003-01-11 & 2003-08-04 & 0.336 & 6.026 & 0.017 & 0.090\\
 2003-08-04 & 2004-04-20 & 0.350 & 5.991 & 0.019 & 0.101\\
 2004-04-20 & 2005-01-13 & 0.350 & 6.025 & 0.021 & 0.100\\
 2005-01-13 & 2006-04-26 & 0.304 & 6.137 & 0.019 & 0.091\\
 2006-04-26 & 2010-12-31 & 0.317 & 6.019 & 0.016 & 0.088\\
\hline
Average&& 0.327 & 6.061 & 0.007 & 0.038\\
\hline
 2002-02-13 & 2010-12-31 & 0.324 & 6.072 & 0.007 & 0.037\\
\hline
\end{tabular}\label{param}
\end{table}

The observed successor of a flare is not necessarily the true one because, e.g. it can be below 
the detection limit or in a phase when the satellite is not active (the instrument is in the 
South Atlantic Anomaly or in the Earth shadow). One may have a concern, therefore, that this 
circumstance inserts some bias on the estimated parameters. For testing the effect of missing 
data on the statistical result, we created a subset of the data by leaving out a large number of 
the observed events.We selected 
only those events that were observed in the 12-25 keV channel and the result can be seen in Figure \ref{12_25-fig}),
which shows that the pattern does not depend on whether the observed successor of a flare is really the true one.
{Thus, we can conclude that the statistical result is not or hardly sensitive to the data missing for any reasons.

The robustness of our approach can be also tested by using the flares in one active region.
We have chosen the active region AR 10162 as a representative sample that contained 253 events 
recorded in the period 17 Oct. 2002 - 31 Oct. 2002, giving us a good basis for statistical investigation. 
Performing the same analysis as before (PCA), we arrive to a similar pattern (see Fig \ref{ar10162}) 
as derived earlier for the full RHESSI database. For this particular active region, we obtained that 
the parameters describing the fitting line are $p=0.3$ and $b=6.2$.

\begin{figure}
\includegraphics[width=84mm]{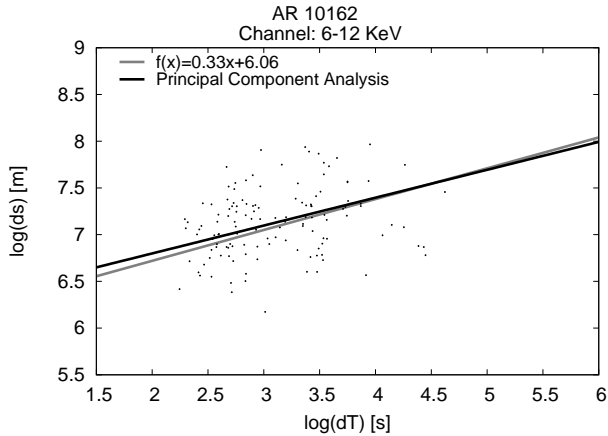}
\caption{Distribution of succeeding flares in the $\{\log_{10}(ds);\log_{10}(dT)\}$ plane in the 6-12 keV 
channel for the active region AR 10162. The meaning of the lines is the same as in Figure \ref{6_12-fig}.}
\label{ar10162}
\end{figure}

\section{Discussion on the physical interpretations}
\label{disc}

\subsection{Sedov-Taylor blast wave}

Table \ref{param} shows that the mean value of the parameter $p$ (the power of the relationship between $R$ 
and $t$) resulting from the fitting line shown in Figures \ref{6_12-fig} and \ref{12_25-fig} is $a=p=0.327\pm 0.007$. 
Based on this finding we will try to explain the relationship between the spatial and temporal differences between 
events assuming that it reflects a physical relationship. 

The first thing we can notice is that of all the $p$ values determined above are close to the value 
of $p\sim1/3$ typical for blast waves in the post Sedov-Taylor phase. The fast-mode freely propagating blast waves 
are a common feature of many flare models, and the reconnection model, that stays at the core of flaring processes, describes two additional types of shocks 
(slow-mode Petschek shocks and fast-mode termination shock), which can have an impact on the ambient medium 
(see, e.g. \citealt{F88}).

The energy release of flares would always act as a temporary piston for a shock that propagates later as
a pressure-driven blast wave (see the review by \citealt{VC08}). Shocks can be also ignited by a smaller-scale process associated with the flare energy release, e.g. expansion of hot loops or small-scale ejecta. 
Numerical simulations show that other alternative mechanisms may be available;
e.g. the downward reconnection jet 
(which may or may not be supersonic) might create a large-scale fast-mode shock wave in the ambient corona after coalescing 
with the law-lying loops and deforming them \citep{Barta08}.

CMEs can also create coronal shocks but this type of shocks propagates in a different way. 
The magnetically driven CME creates a shock which is a combination of the bow-shock and piston-shock, 
and the wave is permanently supplied by the energy from the CME. These CME-driven shocks can later degenerate into large-scale global EIT/EIV waves (see, e.g. Ballai et al. 2005) that can generate oscillations of various remote magnetic structures.
Although blast waves at the onset of energetic flares are theoretically predicted, in most of the observations 
the coronal shock is related to CMEs or CME/flare events. So far there have been found only a few events that demonstrate that the coronal shock wave can be generated by a flare without the presence of an 
associated CME  \citep{M12, KI13}.
In addition, it is most likely that real shock waves are neither purely blast waves nor purely piston
ones \citep{Gea11}. 
Some further considerations may arise when we take into account the favored place of formation of blast wave.
Both theoretical models and observational results are in agreement in that only flares extending 
to the active region periphery are likely to ignite large-scale blast wave because strong field regions are 
unfavorable for generation of flare-associated pressure pulses \citep{VC08}. However, our results could only
be explained in terms of blast waves if there were observational evidences for the 
existence of such flare-ignited small-scale disturbances which can propagate within the active regions as a 
blast wave at least in a part of their propagation time. 
Despite the lack of such observational evidences, it is reasonable to investigate 
whether our results support or refute the existence of such blast waves that 
could cause the determined statistical relationship.

The process of explosive energy release can generate a radially expanding shock wave in a homogeneous 
surrounding medium described by the Sedov-Taylor theory \citep {b33, b34}. After the explosion, the 
shock wave enters into an the adiabatic expansion phase corresponding to a power law of the form given by Eq. (\ref{ST}) with $p=0.4$. 
The coefficient $A$ is related to the ratio of the inputed energy, $E$, of the 
blast wave and the ambient density, $\rho$, via the equation
\begin{equation}\label{Acoef}
A=\left(\frac{E}{\rho B(\gamma)}\right)^\frac{1}{5},
\end{equation}
\noindent where the constant $B(\gamma)$ depends only on the
specific heat ratio, $\gamma$. In the case of $\gamma=5/3$
one obtains $B(\gamma)\approx1$.

Later on the shock wave enters the post Sedov-Taylor expansion phase, when the characteristic time of 
the energy loss is comparable to the characteristic time of expansion of the shell. 
The pressure of hot shocked gas can still drive a shock into the surrounding medium, but radiation 
losses become important. When the shell has expanded and decelerated sufficiently so that its cooling 
time is shorter than its spreading time, it loses energy rapidly by cooling. 
Due to the gradual loss of the intrinsic energy of the wave, the power $p$ drops down to about 1/3. 
In the stage of pressure-driven snowplow (PDS), the analytic value of $p$ is 2/7= 
0.286 \citep{MO77}. The numerical simulations of the transition of the Sedov-Taylor blast wave from 
adiabatic to PDS derive somewhat higher values of $p$ varying in time and depending on the 
numerous characteristics of the ambient medium. 
For example, the power $p$ is expected to be about 0.31 (0.300-0.320) by \citet{b35} or 
about 0.33 (0.312-0.342) by \citet{B98} in the case of thin-shell radiative blast wave.
The final stage is a momentum-conserving phase corresponding to a power
$p=0.25$ \citep{b6}. 

Our statistical method suggests that the relationship between successive flares could follow the 
propagation characteristic of a shock wave in post Sedov-Taylor phase. The result is independent of 
the fact that we see only the projected distance of two spatially separated events onto the plane 
perpendicular to the line of sight. From statistical point of view, the events distributed uniformly 
on a sphere of radius $R(t)$ populate, as a projection, a ring with approximately the same outer radius.
Consequently, the relationship between the time and distance of two succeeding flares remains unchanged. 
Thus, our statistical results may allow us to reveal the basic properties of the determined relationship 
assuming that it describes a physically existent blast wave.

\subsection{Parameter estimation for the blast wave}

The constants $b$ derived by means of the PCA regression impose a constraint upon the value of $A$. 
The fitting with the help of Eqs. (\ref{eqds})-(\ref{eqdT}) results in a range of values 
of $b=\log_{10}(A)= 6.02-6.10$, i.e. $b=6.06\pm 0.04$ as given in Table \ref{param}.

Computing the time derivative of the shock front radius, $R(t)$, we obtain 
the velocity of the propagating shock pattern as 
\begin{equation}\label{vel}
v_R=\frac{dR(t)}{dt}=p A t^{p-1}=p\frac{R(t)}{t}.
\end{equation}
Estimating the inputed energy, the density of the various layers of the solar atmosphere 
and the local magnetoacoustic speed, we can investigate whether a blast wave with the 
obtained values of $A$ can propagate in that medium.

Flares could occur in the solar corona, and it is known that the magnetic reconnection, 
which stays at the core of flare generation, can drive fast shock waves in their surrounding. 
at first we discuss the 
simplest and the most obvious scenario to explain the connectivity of events is 
that the blast wave is generated at the location of the first flare and it propagates on the 
disk in the solar corona until it triggers a subsequent reconnection in the same active region.

Considering the typical density in a coronal active region of the order of $10^{-12}-10^{-13}$ kg m$^{-3}$ 
\citep{b43} and using the value of $b=\log_{10}(A)$, we can derive a range of energies of $10^{17}-10^{19} \, J$. 
We have to note that this estimation has a certain level of uncertainty that could result from the estimations 
of density and constant $A$. The estimation of the density of the ambient medium is not an easy task. 
The density is measured from emission lines, but when looking at the emission in a certain wavelength one 
has to consider the effect of all origins of emissions in the line of sight, therefore estimates of 
densities might be overestimated. The constant $B(\gamma)$ is not affecting the determined energy values 
as its value depending on the specific heat varies within a very narrow interval.

We should note that this energy is not necessarily the entire energy released by the flare and not even 
the whole resulting kinetic energy. The thermal and kinetic energies are not necessarily equal, and the 
equipartition rule between them is not known. Thus, the thermal energy of flares can be used as an 
estimated upper limit of energy of the shock wave. Taking into account that the thermal energy of 
microflares is $10^{19}-10^{23} \, J$, we can estimate that a microflare usually has enough 
energy to generate a shock wave with $A\sim 10^{6.02}-10^{6.10}$ in the solar corona but hardly 
or not at all in the denser lower atmosphere.
Now the question is whether such a shock wave can propagate in the solar corona.

The uncertainty in calculating the coefficient $A$ does not affect the estimation of the velocity 
of the shock pattern. One can see from Eq. (\ref{vel}) that the right hand side 
does not depend on the constant $A$ and consequently on the uncertainties mentioned above.

For calculating the range of propagation speeds, $v_R$, we need the value of the time parameter, $t$, appearing in 
Eq. (\ref{vel}). However, this parameter is a hidden variable in Eqs. (\ref{eqds}) and (\ref{eqdT}) 
due to noise and is not accessible directly from observations, nevertheless it can be calculated 
from the observed $dT$ time differences.

Inspecting the figures listed in the Appendix, one can obtain an estimate for the observed time difference 
to be in the range $2<\log_{10}(dT)<5.5$. According to the PCA method, the hidden time variable, $t$, 
accounts for about 0.7 of the variability of the observed value. Consequently, the range of the 
parameter $\log_{10}(t)$ is $2.5<\log_{10}(t)<5$.

Converting these values to those of the velocity of the shock we obtain a range 
of 0.13 km s$^{-1} < v_R <$  6.7 km s$^{-1}$. If the connection between events is realised 
by a propagating shock wave, this can only propagate in a region where the propagation speed 
of the shock front exceeds the local speed of linear fast magnetoacoustic waves. 
This speed is of the order of several hundred km s$^{-1}$ in the solar corona and it is about 
10 km s$^{-1}$ in the photosphere. This means that the derived speed is completely 
unrealistic for a shock wave propagating in the solar corona or and even in the photosphere. 
Consequently, we obtain a contradiction by assuming a direct blast wave connecting 
two succeeding flare events; therefore, alternative explanations that could shed light on 
the very low propagation speed are needed.

We have three possible ways which can be investigated whether they help to resolve 
this problem: 1) the parameter estimation for the Sedov-Taylor equation needs improvement; 
2) the blast wave propagates only in a fraction of the observed time; 
3) the real physical connection between flares is not a blast wave.

The first possible explanation for the above contradiction is that the direct blast wave 
propagates in the corona but its estimated speed is not correct. This could arise because our 
model lacks the influence of the magnetic field, which may preclude the correct parameter 
estimation. In an active region, the dynamics is driven by magnetic forces which 
can alter Eq. (\ref{Acoef}).
However, this simplification cannot explain why the derived propagation speed is smaller 
by several order of magnitude than the local speed of linear fast magnetoacoustic waves in the corona.

The second possible way to resolve the problem is to assume that the blast wave is only 
a part of a sequence of the dynamic events generated by the instigator flare. 
If the time for the disturbance to propagate in form of a blast wave is only a fraction 
of the elapsed time we observed between two succeeding flares, the propagation speed of the 
blast wave can be larger than the speed determined for the whole time.
However, one order of magnitude increase in speed of blast wave would need five orders of 
magnitude increase in energy of flare. Since the speed would need about four orders of 
magnitude increase, thus we have to conclude that the microflares probably cannot generate such a 
Sedov-Taylor blast wave propagating in the solar corona that could explain the results presented in this paper.

If we assume that the disturbance propagates in form of a blast wave only in a fraction of both the 
total spatial and temporal distances, it may formally resolve the problem. 
The available energy can be enough for a blast wave to propagate a small fraction of the spatial difference
with an appropriately high speed in a fraction of the elapsed time.
However, in this case, the equations would contain more unknown parameters than that could not be 
determined by using the present method because we do not know how the disturbance propagates in 
the remaining part of its propagation and what else happens during the remaining part of the time interval. 
We can only discuss the known models, observations, 
and simulations whether they allow such a scenario.

According the standard flare model, the flare taking place in the corona can generate energetic 
disturbances that can travel towards the footpoints in the denser lower atmosphere. 
These disturbances arriving in the chromosphere or the photosphere can trigger blast waves. 
If we assume that such a type of blast waves plays a role in the found relationship, it allows 
the blast wave to propagate only a part of the observed distance in a part of the observed time.
It is reasonable to assume that a blast wave propagating in the photosphere can collide with 
a neighbouring flux tube triggering a disturbance propagating upwards, eventually leading to a 
succeeding flare. In this scenario, the observed time consists of three different time intervals. 
The first one is the travel time of transporting the energy from the location of the flare to 
the triggering site of the blast wave. The second one is the time the blast wave spends in-between 
its origin and the footpoint of the neighbouring flux tube and finally, the time necessary for 
a secondary disturbance to reach the location of the succeeding flare event from the place 
where the blast wave collided with the magnetic flux tube.

The triggering disturbance for the subsequent flare is probably a plasma upward flow 
according to the simulation by \citet {Sea00}.
These authors studied the process of interaction between shock waves and magnetic flux 
tubes taking into account the effect of the gravitationally stratified background density, 
and they found that a strong upward plasma jet, as well as surface Alfv\'en waves can 
propagate along the flux tube. The shock wave travelling to a current sheet may initiate 
magnetic reconnection process according to the simulations by \citet {OK97}, which supports 
the idea that the solar flare can be triggered by the shock wave coming from a distant flare.

This scenario is also supported by the observation of \citet {Mea03}. They found a triggering 
strong plasma upward flow before a B-class flare and after a few minutes, an impulsive event 
was detected at chromospheric heights. About 10 minutes later, they observed downward plasma 
pulses with energies of the order of $10^{18}-10^{19} \, J$ penetrating down to the photospheric 
layers, and they observed eleven similar downflows during 47 minutes all together. The duration 
of the impulses was always of the order of 5–10 minutes, and they caused shock waves in the 
photosphere with an initial horizontal propagation speed of about 20 km s$^{-1}$.
By using the real energy of flares, we can estimate the range of $log_{10}(A)$ to be 
between 4.8 and 5.4 in similar cases. 
If the $log_{10}(A)$ is in this range, the shock front can be estimated to propagate 
horizontally a distance of about 100-600 km within a few seconds. 
If the shock wave collides with a flux tube during this propagation, this may result 
in a flare later according to the above mentioned simulations by \citet {Sea00} and \citet {OK97}.  
Based on these results, it seems possible that a flare-generated shock wave can contribute 
to the generation of a next flare in such a way. 
However, it cannot be decided now whether the outlined scenario works or not.

\subsection{Other possible physical interpretations}

It is also possible that the connection between the succeeding flares is not even realised 
by a shock wave but by an unknown physical mechanisms, whose kinematic characteristics 
resemble the properties of a blast wave. 
In a recent paper, \citet {a12} investigated 155 M- and X-class flare events seen by GOES 
and SDO/AIA satellites and measured the spatial and temporal expansion of flare areas. 
His main finding was that for the majority of the cases the expansion was sub-diffusive with 
the expansion rule obeying the relation $r(t)\sim t^{\beta/2}$ where $\beta=0.53\pm 0.27$ 
that could be attributed to the an anisotropic chain reactions of intermittent magnetic 
reconnection episodes in a low plasma-beta corona. The speed of area propagation was of 
the order of $15\pm 12$ km s$^{-1}$, a value that cannot be explained in terms of MHD waves. 
Despite the extremely large error resulted in his statistics ($\approx 50\%$) the value of 
$\beta/2$ is very close to the value we obtained by fitting a much larger sample of RHESSI data. 
Based on this model, we can imagine that the area of a flare in its expansion triggers subsequent flares in the same AR.

During a flare, the change of the magnetic structure in the corona is reflected in the
motion of chromospheric H$\alpha$ flare ribbons and corresponding HXR footpoint sources.
It is often observed by RHESSI that the source of HXR flux has a 
spatial displacement along the arcade of magnetic loops during a flare. 
This could be caused by some disturbance propagating along the arcade, sequentially triggering 
a reconnection process in successive loops of the arcade (e.g. \citealt {GB05, Yea09}). These 
consecutive flaring events identified from a single flare may be observed as successive flares 
as they produce separate HXR peaks. The mean velocity of the footpoint motion determined by 
\citet{b41} is about 5-70 km s$^{-1}$ which is of the same order of magnitude as the propagation speed derived 
from our results in the time range between a few seconds and a few minutes. Thus, it is reasonable 
to assume that the above-mentioned features may contribute to the found statistical relationship 
in the time range of duration of flares.
On the times scales of hours and days, the evolution of active regions probably has a large 
contribution to the change of the position of flaring sites or dominates it.

Finally, our results could be a signature of a much intricate physics connected to the 
reorganisation of the magnetic field following a flare. Flares always occur at the boundary 
of large neighbouring patches with opposite vertical magnetic field. When these fields are 
pushed together, reconnection takes place and a huge amount of energy is released. 
Flares are observed to take place where the magnetic field has a strong distorted 
configuration (twist of shear), meaning that the only way to release the magnetic helicity 
stored in magnetic field lines is through the eruption. These events are followed by a 
large-scale reorganisation of the magnetic field whose kinematic parameters
would follow the relationship we found earlier.

\section{Summary and conclusions}
\label{sum} In the present paper we studied the statistical relationship 
between succeeding flare events recorded by the RHESSI satellite. 
We selected the flares that were associated to active regions in the period 2002-2010.

The heliographic coordinates of events were calculated applying the appropriate corrections.
With the help of the coordinates we were able to determine spatial distances between 
succeeding events ($ds$) in the same active regions, while temporal differences 
($dT$) were deducted from the RHESSI catalog.

We studied the statistical relationship between the $ds$ and $dT$ differences 
dividing the RHESSI data in subsamples according to the phase of the solar cycle 
and the energy channels, respectively.

If we take into account only flares occurring in the same active region, we obtain 
a linear relationship in the $\{\log_{10}(ds)-\log_{10}(dT)\}$ diagram indicating 
a power law connection between the spatial and temporal differences of succeeding flares.

This relationship may reveal a hidden factor responsible for the statistical 
connection. Performing a Principal Component Analysis (PCA) on the $ds$ and $dT$ 
values we obtained the parameters of the power law relationship. The parameters 
obtained in this way showed a remarkable homogeneity, independently from the 
phase of the solar cycle.

Several possibilities were discussed to explain the temporal and spatial correlation 
between events. One candidate that could explain the connectivity was a blast wave. 
This assumption was made based on the statistically determined time-distance relationship 
of the form $R(t)=At^p$, where $p=0.327\pm 0.007$, i.e. very close to the typical value of a 
shock in the post Sedov-Taylor stage. The most problematic point of 
this scenario is the derived propagation speed of these shocks being sub-sonic, 
at least a few orders of magnitude less than expected. We had to assume that the 
blast wave transmitting the causal connection between two succeeding flares takes 
only a few percent of their spatial and temporal differences if it plays a role at all.

It seems more possible, that the net effect of different, basically magnetic, processes in 
the solar atmosphere mimics statistically a functional relationship between the time and 
spatial distances of succeeding flares in a mathematical form of a blast wave. 
In particular, the relationship between temporal and spatial distances we observed 
could be the characteristics of the magnetic field re-arrangement after a flare 
in the same active region.

Finally, we need to note that our derived values are similar to the values obtained 
earlier by \citet {a12} using SDO data, where the spatial and temporal differences 
were connected to the expansion of the active region area in the sub-diffusive stage. 
If our data analysis would point to the same mechanism described by this author, 
it would be a further evidence for the universal character of it as, according to 
our findings, this occurs not only in particular active regions but it can be shown 
to be a trend over much longer periods.

\section*{Acknowledgements}
This work was supported by the OTKA grants No. K77795, K81421 and K83133.
The research leading to these results has received funding from
the European Community's Seventh Framework Programme
(FP7/2007-2013) under grant agreement eHEROES (project No.
284461).

\appendix
\section[]{Spatio-temporal distribution of succeeding flares}

In this Appendix we show a collection of scatter plots representing the distribution 
of the succeeding flares in the $\{\log_{10}(ds);\log_{10}(dT)\}$ planes observed in the 6-12 keV energy 
channel and in different phases of the solar cycle.  

The large scatter of the data can be explained by the observed spatial distribution of flares. 
It has long been known that solar flares tend to occur along magnetic polarity inversion lines (PIL), which
are places favourable for repeated flaring (e.g. \citealt{H84}). If there is one PIL in an active region, this 
means that the most probable position of the next flare close in time is in the vicinity of the previous one. 
If there are several PILs in the active region, the next flare may happen at a different PIL, thus the spatial 
distances can have a large scatter in these cases. In our large statistics of flares, these two facts will
cause a log-normal distribution of spatial distances. This explains that the $\log_{10}(ds)$ data have a
large scatter ranging from the spatial resolution of the RHESSI observations to the size of the large sunspot groups.
The Shapiro-Wilk test shows that  we can accept the assumption of normal distribution of $\log_{10}(ds)$ in the following intervals of $\log_{10}(dT)$: 2-2.5, 2.5-3, 3-3.5, 3.5-4. This supports the assumption of log-normal distribution of spatial distances in the case of time difference smaller than about 2-3 hours.  In the case of larger time differences,
the evolution of sunspot groups seems to cause such changes in the position of places favourable for flaring that the
probability of small spatial differences decreases. This somewhat decreases the scatter of data points 
but it results in the rejection of the normal distribution of $\log_{10}(ds)$.

Because of this large scatter of the data, the PCA regression gives a more 
reliable result than the ordinary least squares method. 
By definition the first principal component (PC)
represents the maximum variance among the linear combinations
of the observed variables. This PC defines a direction which
gives correctly the trend and its slope of the point pattern
in the plain of the observed variables. In contrast, the
ordinary least squares method systematically underestimates the
value of this slope if both the observed
variables contain a stochastic noise term (see, e.g. \citealt{D68} for the study of the role of the PCA in astronomical context).

We have made a compromise between the resolution according to the phase of the 
solar activity cycle and the accuracy of determining the statistical parameters 
of the subsamples. To ensure the same numbers of points (1800) in these scatter 
plots, the time intervals corresponding to the different figures are different. 
Note the remarkable homogeneity of the distributions independently of time.
The chi-square test based on the data in Table 2 shows that the variations of $a$ and $b$
are not significant. The result of chi-square test is similar if the data set is dividied
into six subintervals with equal length of one and a half year. This means that 
the parameters $a$ and $b$ can be reckoned as constants.
This homogeneity of the parameters of the fitted lines makes serious restrictions 
for the models trying to provide with a theoretical explanation: the parameters 
responsible for this effect should be independent of the actual phase 
of the solar activity cycle.

\begin{figure*}
\begin{centering}
\includegraphics[width=80mm]{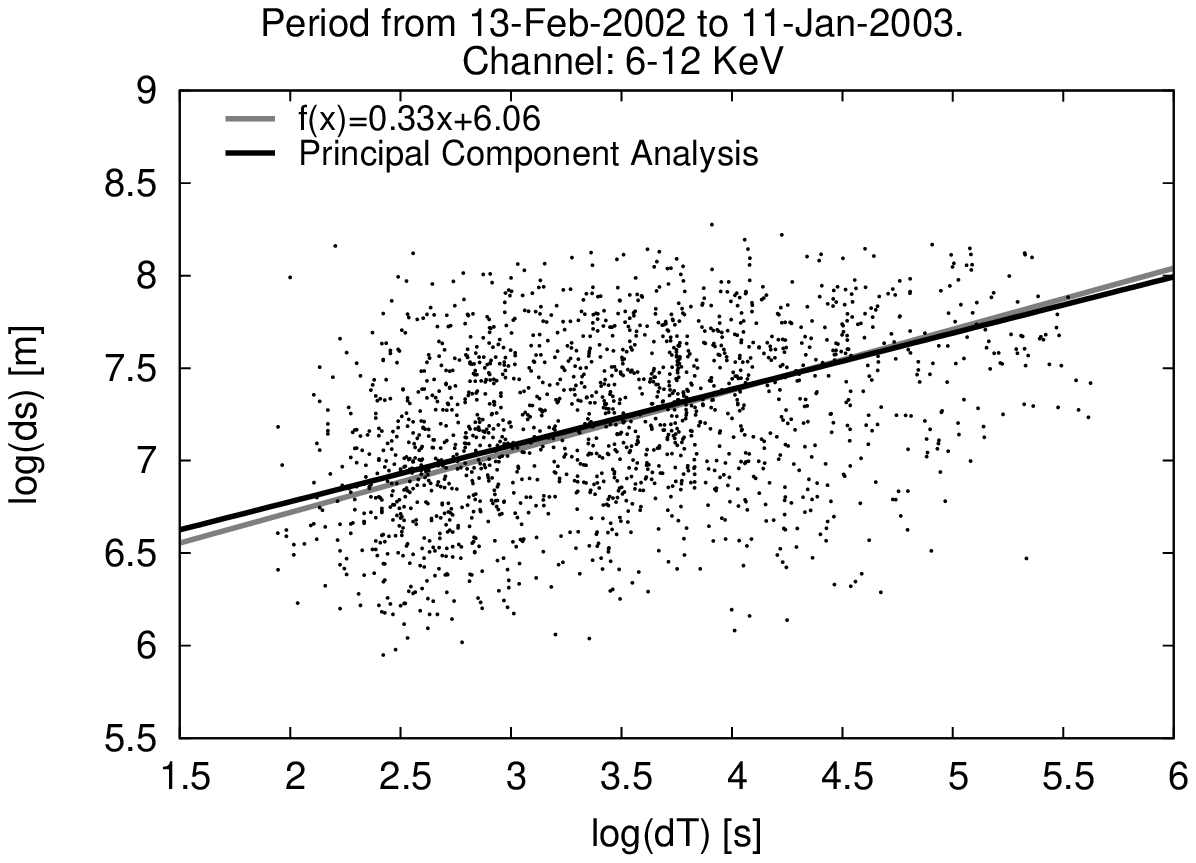}
\includegraphics[width=80mm]{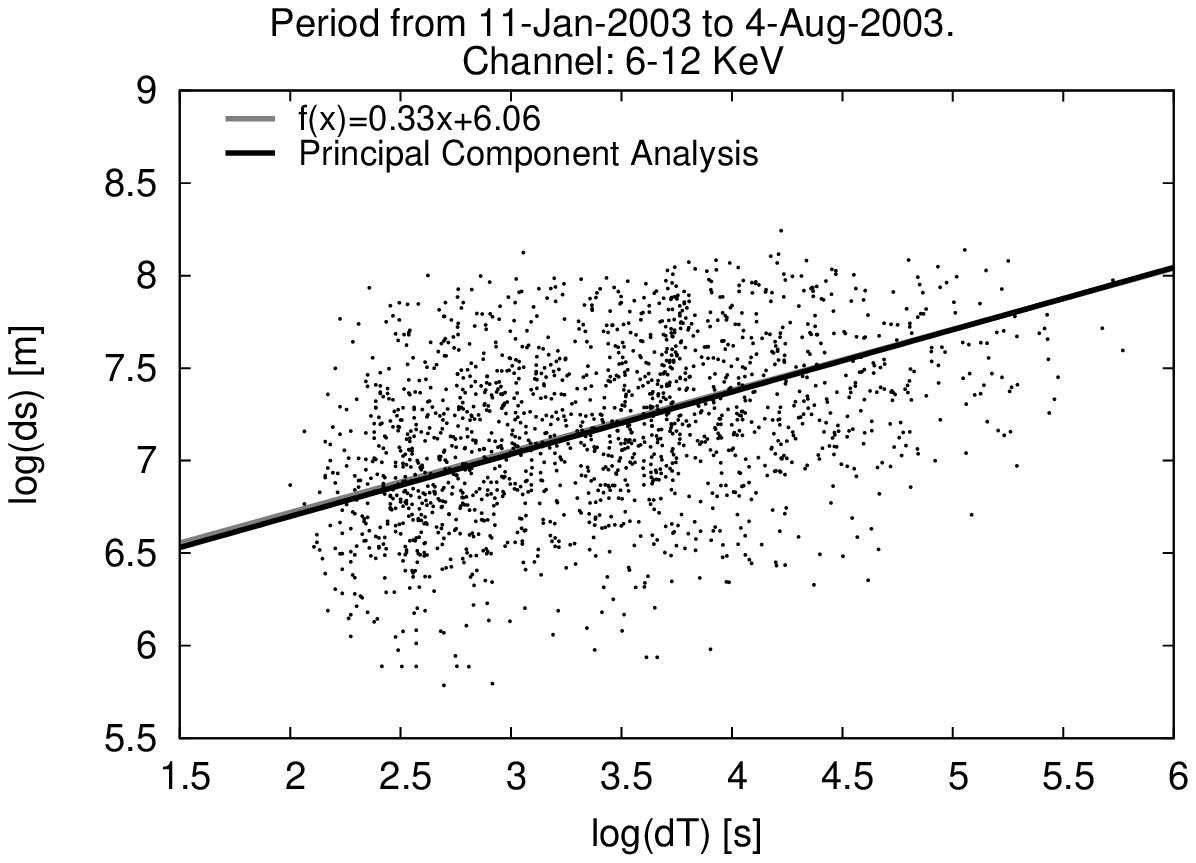}
\label{Fig_app_1_2}
\end{centering}
\end{figure*}

\begin{figure*}
\begin{centering}
\includegraphics[width=80mm]{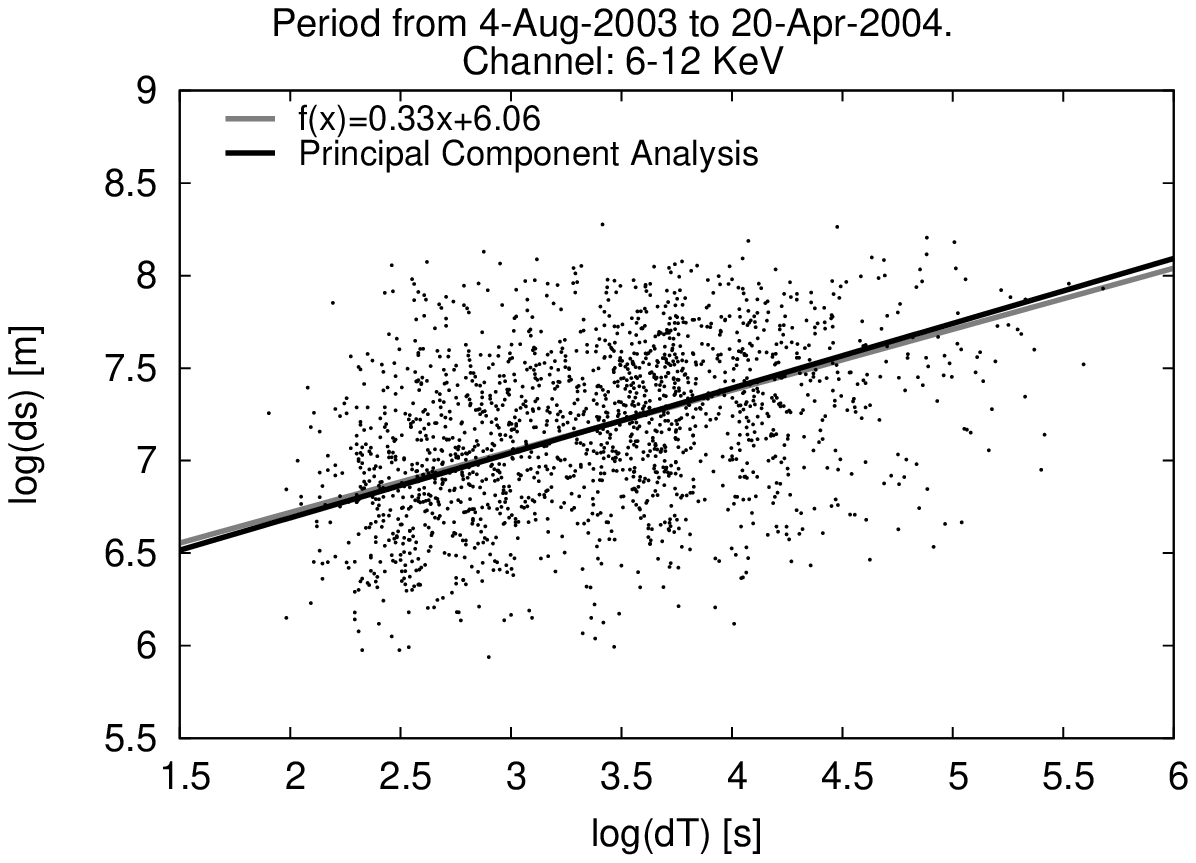}
\includegraphics[width=80mm]{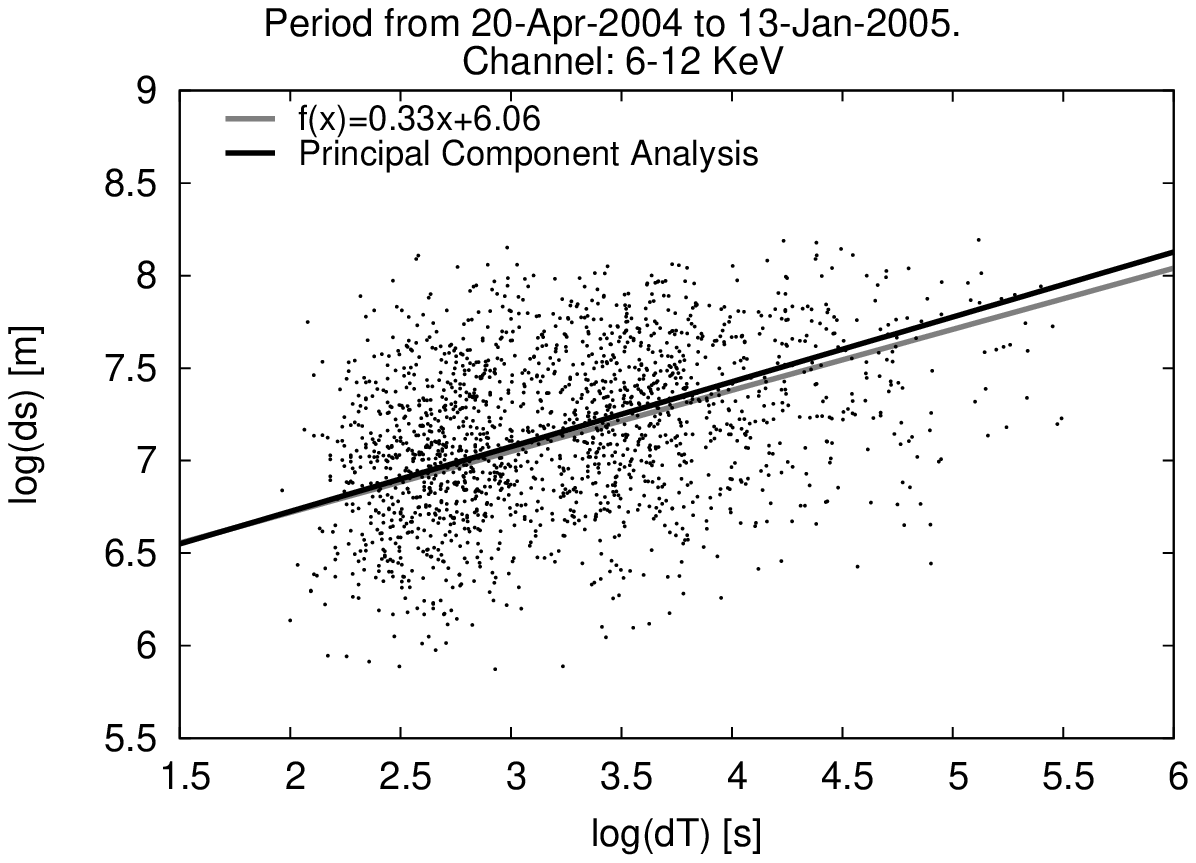}
\label{Fig_app_3_4}
\end{centering}
\end{figure*}

\begin{figure*}
\includegraphics[width=80mm]{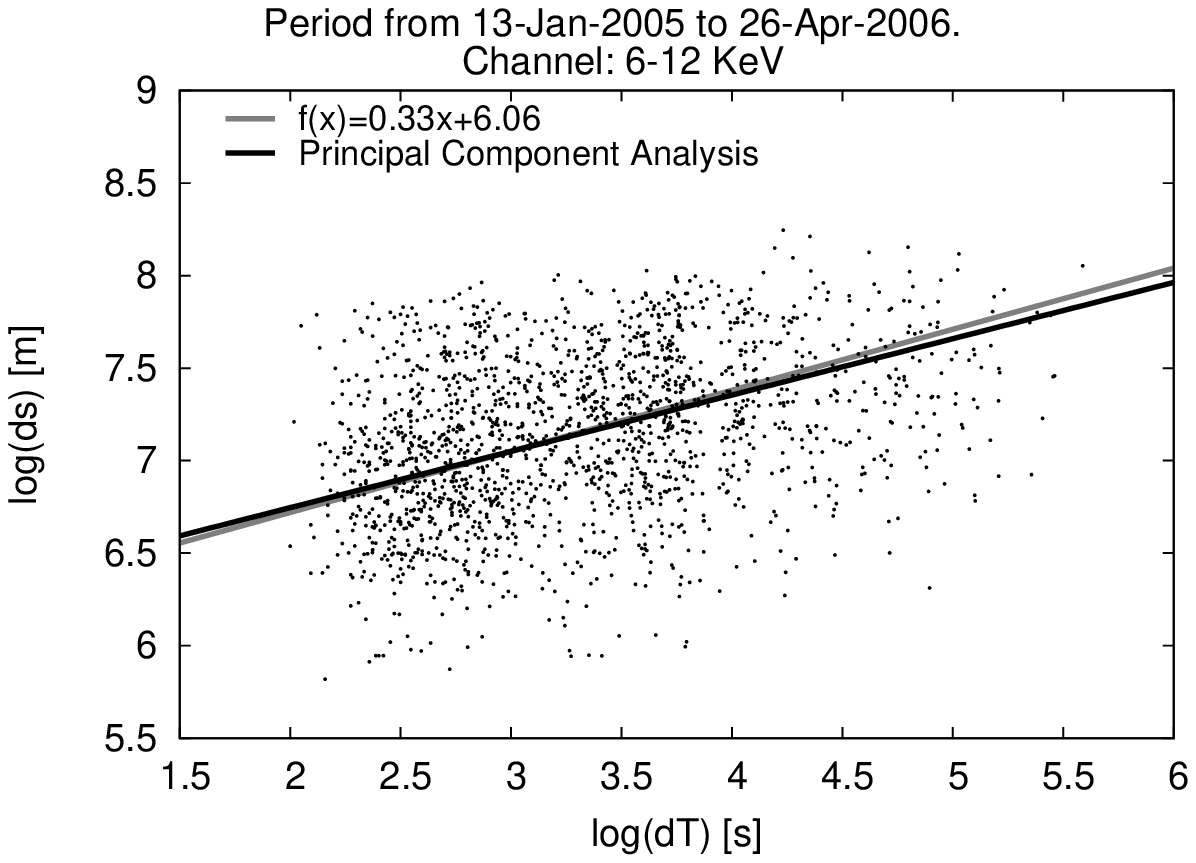}
\includegraphics[width=80mm]{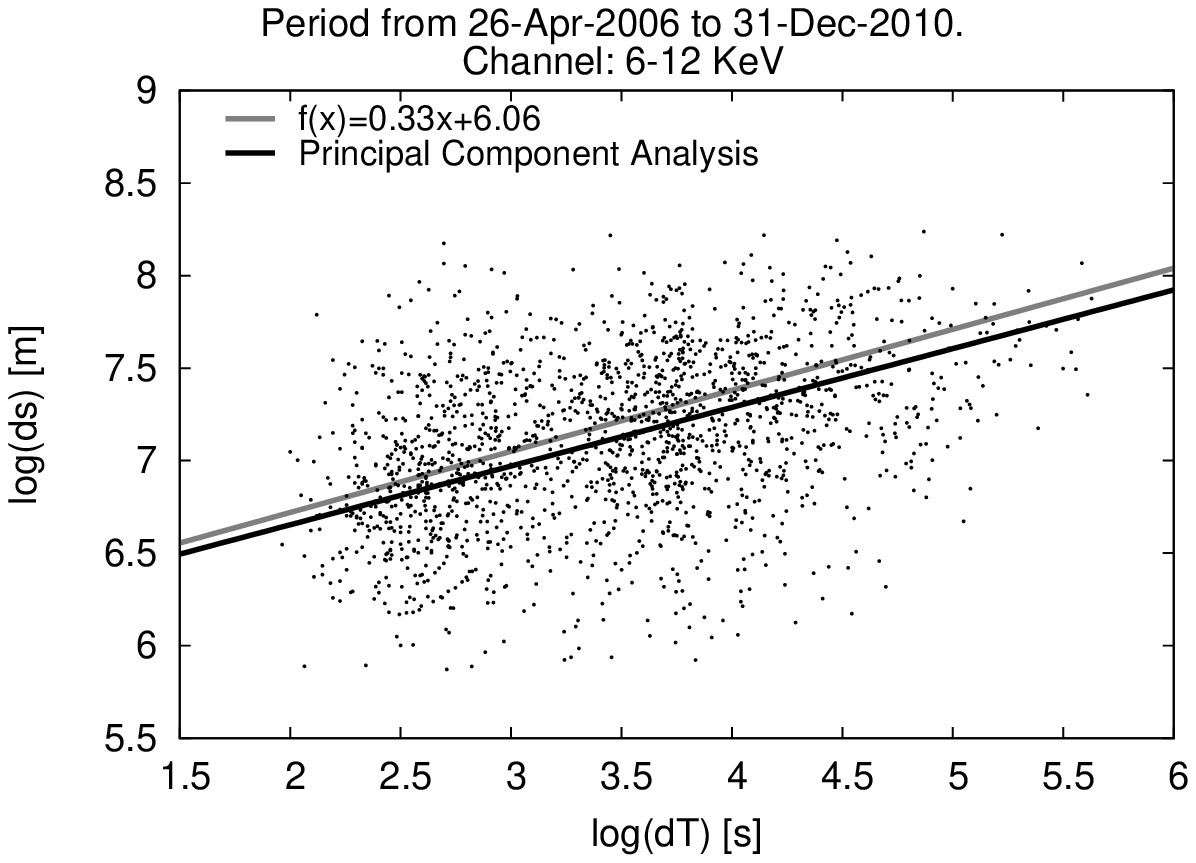}
\label{Fig_app_5_6}
\caption{Distribution of flares measured in the 6-12 keV channel and 
their successors in the $\{\log_{10}(ds);\log_{10}(dT)\}$ plane. 
The black solid line indicates the actual result of the PCA regression 
and the grey line is determined by the average parameters in 
Table \ref{param}.}
\end{figure*}

\label{lastpage}
\end{document}